\documentclass[twocolumn,preprintnumbers,aps,amsmath,amssymb]{revtex4}

\usepackage{graphicx}
\usepackage{dcolumn}
\usepackage{bm}
\usepackage{epsfig}

\begin{document}


\title{A note on Kauzmann's paradox}

\author{U. Buchenau}
\affiliation{%
Forschungszentrum J\"ulich GmbH, J\"ulich Centre for Neutron Science (JCNS-1) and Institute of Biological Information Processing (IBI-8), 52425 J\"ulich, Germany}%

\date{October 11, 2022}

\begin{abstract}
The rapid structural and vibrational entropy decrease with decreasing temperature in undercooled liquids is explained in terms of the disappearance of local structural instabilities, which freeze in at the glass temperature as boson peak modes and low temperature tunneling states. At the Kauzmann temperature, their density extrapolates to zero, as evidenced by neutron and heat capacity data in liquid selenium.
\end{abstract}

\maketitle

In 1948, Walter Kauzmann \cite{kauz} realized that the excess entropy $S_{exc}$ of an undercooled liquid over the crystal extrapolates to zero at the nonzero Kauzmann temperature $T_K$. Kauzmann's paradox lies in the difficulty of finding a theory for the disappearance of the structural entropy of the liquid at a finite temperature \cite{cavagna,berthier}. 

Of course, the undercooled liquid undergoes a kinetic freezing into the glass phase at the glass transition temperature $T_g>T_K$, thus avoiding Kauzmann's paradox. But in fragile glass formers, $T_K$ lies closely below $T_g$, suggesting a true thermodynamic second order phase transition \cite{cavagna,berthier} at $T_K$. For simple liquids consisting of atoms interacting with pair potentials, there is a mean field theory (Random First Order Theory) predicting a nonzero Kauzmann temperature \cite{parisi}, but the influence of fluctuations seems to bring the Kauzmann temperature down to zero in two dimensions \cite{berthier}.

Later, it became clear that not only the excess entropy extrapolates to zero at the Kauzmann temperature, but also two other important quantities, namely the inverse viscosity \cite{as,ra,angell} and the excess mean square displacement \cite{zorn,bzr,hansen1,hansen2} of the undercooled liquid over the crystal. The latter disappearance is due to a strong reduction of the number of soft modes responsible for the boson peak and the tunneling states in glasses with decreasing temperature in the undercooled liquid, evidenced both by experiment \cite{se,hansen1,hansen2} and numerical work \cite{wang,bertun} using the new swap mechanism \cite{swap}.

The soft modes in glasses, responsible for the low temperature glass anomalies, are well described by the pragmatical soft potential model \cite{ramos,herbert,abs}, an extension of the tunneling model \cite{phillips} to include soft vibrations and low barrier relaxations. New numerical results (for a review see \cite{bouch}) confirm three soft potential model predictions, namely the existence of a strong fourth order term in the mode potential \cite{le1}, a density of soft vibrational modes increasing with the fourth power of their frequency \cite{wang,le1,manning,le2,mizuno}, and a density of low barrier relaxations increasing proportional to the power 1/4 of the barrier height \cite{le4}. One should note, however, that other models for the boson peak exist \cite{schirmacher,grigera}, and that some numerical results \cite{angelani,paoluzzi} question the fourth-power density of states. 

Another crucial numerical finding \cite{corei,corein} is the unstable core of the soft modes, stabilized by the surroundings. The unstable core of the soft mode is a local saddle point configuration between two minima, lower than the saddle point in energy by the creation energy $E_c$ of the soft mode. $E_c$ was found to be about 2.5 $k_BT_g$ in a binary glass former \cite{edan}. Since the modes couple strongly to an external shear \cite{edan,le6}, the core is a local shear instability, and the short time shear modulus decreases with an increasing number of soft modes. This explains the strong decrease of the shear modulus with increasing temperature \cite{nemilov,dyre}.

The soft mode concentration at a given temperature and density in the undercooled liquid is determined by its free energy $E_f=E_c-TS_c$, where $S_c$ is the vibrational entropy \cite{se,wyart} (a softer mode has a larger number of occupied vibrational levels between which it can choose). The increase of the vibrational entropy has been analyzed in detail from inelastic neutron scattering data in selenium \cite{se}. It was found that it is a fraction 0.3 of the total entropy increase, having the same temperature dependence as the total. Selenium is a short chain polymer \cite{sen} with one anharmonic Van-der-Waals, one bond bending, and one bond stretching degree of freedom per atom. The present paper shows that one can find a new answer to the old Kauzmann riddle in this heavily studied glass former.

\begin{figure}[b]
\hspace{-0cm} \vspace{0cm} \epsfig{file=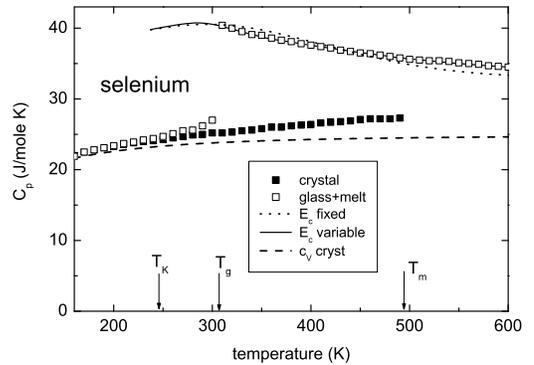,width=7 cm,angle=0} \vspace{0cm} \caption{Heat capacities of crystalline, glassy and liquid selenium \cite{wunderlich} compared to the present calculations for the liquid (continuous+dotted line). The dashed line is the harmonic expectation for the crystalline vibrational density of states.}
\end{figure}

The heat capacity data of crystalline, glassy and liquid selenium \cite{wunderlich} are shown in Fig. 1. The crystalline value increases above the Dulong-Petit expectation for a perfectly harmonic solid by about 0.3 $k_B$ per atom at the melting temperature, demonstrating a visible influence of the anharmonicity even in crystalline selenium.

In order to separate harmonic and anharmonic influences on the liquid heat capacity, one can calculate the difference
\begin{equation}\label{cpcv}
	c_p-c_V=TB_lV_m\alpha_l^2
\end{equation}
from the molar volume $V_m=18.47$ $10^{-6}$ m$^3$, the thermal volume expansion $\alpha_l=3.63$ $10^{-4}$ K$^{-1}$, and the liquid bulk modulus $B_l=6.75$ GPa \cite{simha} at $T_g=307$ K. One finds that the anharmonic $c_p-c_V$ explains about one third of the difference $\Delta c_p$ between liquid and crystal. The rest is close to the theoretical limit of 1/2 $k_B$ for a structural strain energy in each vibrational degree of freedom, not only in the anharmonic Van-der-Waals ones, but also in the harmonic bond bending and bond stretching. 

At first sight, the anharmonic component of eq. (\ref{cpcv}) increases linearly with temperature. In fact, the thermal expansion $\alpha_l$ remains constant up to temperatures above the melting point \cite{campbell}. But the liquid bulk modulus, dominated by the weak van-der-Waals interaction between selenium atoms in neighboring chains, decreases with increasing volume. The average Gr\"uneisen parameter $\gamma$ of both glass and crystal \cite{grosse} at $T_g$ is 1, implying a van-der-Waals Gr\"uneisen parameter $\gamma=3$, which together with the known thermal expansion $V_m\propto T^{0.11}$ implies $B_l\propto T^{-0.33}$. Consequently, $c_p-c_V=(0.61\pm 0.04)(T/T_g)^{0.78}k_B$ per atom: the anharmonic component still increases markedly with temperature.

Since the whole liquid $c_p$ decreases with increasing temperature, it follows that the remaining harmonic structural strain component must decrease strongly with increasing temperature.

To explain this decrease, the present paper postulates a limitation of the structural distortion energy per vibrational degree of freedom by a cutoff energy equal to the free creation energy $E_f$ of the soft modes. For higher energies, the vibrational degree of freedom can get rid of its high energy by the creation of a soft mode in its neighborhood. The postulate implies an error function structural strain energy $E_s$ per degree of freedom
\begin{equation}
	E_s=\frac{1}{2}k_BT{\rm erf}\left(\frac{E_f}{k_BT}\right),
\end{equation}
which with
\begin{equation}
	a=\frac{E_f}{k_BT}\ \ b=\frac{E_f-T\partial E_f/\partial T}{k_BT}
\end{equation}
gives rise to the heat capacity component
\begin{equation}\label{cps}
	c_{ps}=\frac{1}{2}k_B\left({\rm erf}(a)-\frac{2\exp(-a^2)b}{\sqrt{\pi}}\right)
\end{equation}
per degree of freedom.

\begin{figure}[b]
\hspace{-0cm} \vspace{0cm} \epsfig{file=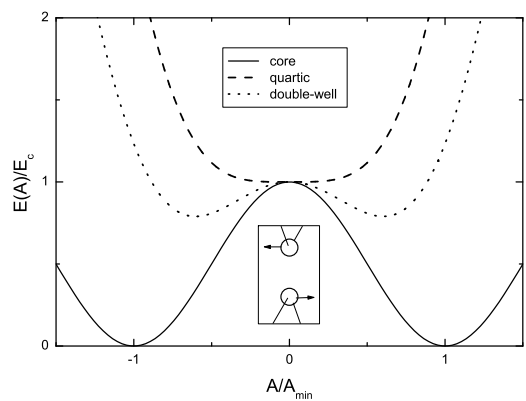,width=7 cm,angle=0} \vspace{0cm} \caption{Mode potentials in units of the creation energy $E_c$ of the soft mode as a function of the normal coordinate $A$: continuous line core potential, dashed line quartic soft mode potential for perfect compensation, dotted line tunneling state potential. The insert shows the opposite motion of two selenium atoms in two compressed neighboring chains, a schematic picture of the unstable core of the mode.}
\end{figure}

In selenium, the local instabilities appear in the van-der-Waals modes \cite{se}, most probably a local shear displacement of two compressed neighboring chains against each other, as indicated schematically in the insert of Fig. 2. The mode is soft because the negative contribution to the restoring force from the unstable core potential (the continuous line in Fig. 2, described here by a cosine function in the normal coordinate $A$ of the mode), is compensated by the positive elastic restoring forces from the surroundings. One gets the purely quartic potential (the dashed line in Fig. 2) for perfect compensation and if the zero point of the harmonic outside forces happens to lie exactly on the saddle point. If the outer forces happen to be weaker or $E_c$ happens to be higher, one gets a low-barrier double-well potential (the dotted line in Fig. 2), such as one needs to explain low temperature tunneling states \cite{phillips}. 

It is clear that in this situation one expects a constant distribution of linear and quadratic terms adding to the purely quartic potential, the central assumption of the soft potential model \cite{ramos,herbert,abs}. These modes freeze in at the glass temperature and provide the boson peak and the tunneling states, which dominate the heat capacity, the thermal conductivity and the sound absorption at low temperatures.

The mean square displacement of the purely quartic potential $Wx_s^4$ with the zero point energy $W$ at the glass temperature $T_g$ is
\begin{equation}
	<x_s^2>=\frac{1}{\sqrt{2}\pi}(\Gamma(3/4))^2\left(\frac{k_BT_g}{W}\right)^{1/2},
\end{equation}
from which one determines an average soft mode frequency $\omega_4$ at the glass transition. The soft potential fit of the low temperature anomalies in selenium \cite{ramos} supplies $W/k_B=1.3$ K, which at the glass temperature of 307 K leads to $\hbar\omega_4=1.07$ meV, a bit lower than the boson peak. This is shown in Fig. 3.

\begin{figure}[b]
\hspace{-0cm} \vspace{0cm} \epsfig{file=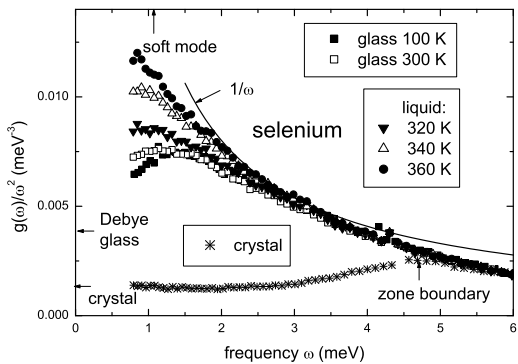,width=7 cm,angle=0} \vspace{0cm} \caption{The quartic soft mode at $T_g=307$ K lies below the boson peak, a factor of four to five lower than the zone boundary frequency in selenium \cite{se}.}
\end{figure}

If the core potential is indeed a cosine, it is straightforward to determine the ratio between the mean square displacements at the saddle point and in one of the minima. The square root of this ratio provides the frequency ratio, which allows to calculate the free energy of the saddle point
\begin{equation}\label{ef}
	E_f=E_c-k_BT(0.675+\ln(E_c/k_BT)/4).
\end{equation}
Eq. (\ref{ef}) contains not only the ratio of the mean square displacements, but also the difference in the average vibrational potential energy, which in the quartic potential is only $k_BT/4$. 

Assuming a constant $E_c=2.9\ k_BT_g$ at all temperatures, one gets the dotted curve in Fig. 1, already providing a reasonable fit of the data. To get the better fit of the continuous line, one has to admit a slight temperature dependence of $E_c$, beginning with 3.1 $k_BT_g$ at $T_g$ and dropping to 2.9 $k_BT_g$ around 400 K. At $T_g$, $\partial \ln{E_c}/\partial \ln{T}\approx0.9$. These two values imply a soft mode concentration increasing with $\exp(-4T_g/T)$ at $T_g$, within the error bars equal to the logarithmic increase 4.4 calculated for quartic soft mode potentials from the measured logarithmic increase \cite{zorn,bzr} $\partial\ln<u^2_{exc}>/\partial\ln{T}=4.9$ of the excess mean square displacement $<u^2_{exc}>$ of liquid selenium over crystalline selenium.

If one identifies the Kauzmann temperature with the one where the density of the soft modes extrapolates to zero, the Kauzmann paradoxon is resolved, because the disappearance at a finite temperature is a simple consequence of the linear down extrapolation of an exponential increase, no longer containing the postulate of a real disappearance of the excess entropy at a finite temperature.

The same argument holds for the Vogel-Fulcher temperature, where the viscosity diverges. In selenium \cite{zorn}, the viscosity obeys the relation $\eta=\eta_0\exp(u_0^2/<u^2_{exc}>)$ over the whole temperature range from $T_g$ to 600 K, providing a better fit than any Vogel-Fulcher law. But its linear extrapolation to lower temperatures is a Vogel-Fulcher law, with a Vogel-Fulcher temperature equal to the Kauzmann temperature. 

The concept of a structural strain energy in each vibrational degree of freedom is supported by the Prigogine-Defay ratio of the glass transition \cite{jackle,schmelzer,speedy}
\begin{equation}\label{prigo}
	\Pi=\frac{\Delta c_p\Delta\kappa}{(\Delta\alpha)^2T_gV_m}=\frac{\overline{\Delta H^2}\ \ \overline{\Delta V^2}}{(\overline{\Delta H\Delta V})^2},
\end{equation}
which relates the increases of the heat capacity at constant pressure $\Delta c_p$, of the compressibility $\Delta\kappa$ and of the thermal volume expansion $\Delta\alpha$ at the glass temperature $T_g$ to the additional enthalpy and volume fluctuations $\Delta H$ and $\Delta V$ in the liquid, respectively. $V_m$ is the molar volume. If the enthalpy and volume fluctuations are completely correlated, the Prigogine-Defay ratio is one. But if there are contributions from different vibrational modes $i$ with different Gr\"uneisen parameters $\gamma_i=\Delta V_i/\kappa\Delta H_i$, then
\begin{equation}\label{gamma}
	\Pi=\frac{\overline{\gamma^2}}{\overline{\gamma}^2},
\end{equation}
with the $\gamma$-values weighted with their contribution to $\Delta c_p$. In selenium, where about half of $\Delta c_p$ is due to harmonic modes, one expects a Prigogine-Defay ratio close to 2, in agreement with experimental results \cite{simha,schmelzer}.

Since the two harmonic covalent degrees of freedom contribute only to $\Delta c_p$ and practically nothing to $\Delta\alpha$ nor to $\Delta\kappa$, one expects (and finds, not only in selenium \cite{simha}, but also in many other glass formers \cite{as}) the validity of the fourth Ehrenfest equation \cite{speedy}
\begin{equation}
	\frac{dT_g}{dp}=\frac{\Delta\kappa}{\Delta\alpha},
\end{equation}
while the first Ehrenfest equation gives a $dT_g/dp$ which is a factor $1/\Pi$ too small.

Selenium is typical for glass formers with a mixture of soft van-der-Waals modes and hard covalent ones. In molecular glass formers, the harmonic hard degrees of freedom are molecular vibrations, in addition to the six van-der-Waals degrees of freedom. In the two examples 2-methyltetrahydrofuran (C$_5$OH$_{10}$) and 3-bromopentane (C$_5$BrH$_{11}$) $\Delta c_p$ corresponds to 8.91 and 9.1 $k_B$ per molecule \cite{ra}, indicating 18 participating degrees of freedom, in the simplest possible interpretation the degrees of freedom of the six non-hydrogen atoms per molecule. The same interpretation works in n-propanol (C$_3$OH$_8$) with a $\Delta c_p$ corresponding to 6.35 $k_B$ per molecule \cite{ra} and in glycerol (C$_3$O$_3$H$_8$) with a $\Delta c_p$ corresponding to 9.6 $k_B$ per molecule \cite{giau}. In this last case, $\Delta c_p$ is as large as the $c_p$ of the glass, showing that a degree of freedom can take up distortion energy without the quantum limitation which holds for the uptake of vibrational energy. There are other even more drastic examples like sulfuric acid with water \cite{wunder}.

A much smaller participation of additional degrees of freedom is found in polymers \cite{wunder}, where $\Delta c_p$ corresponds on the average to 1.4 $k_B$ per "bead". Here every chain atom counts as a bead, so the polystyrene monomer C$_8$H$_8$ consists of two beads, where half of the beads carry a phenylene ring. But these phenylene rings do not seem to contribute; polystyrene is in the middle of the polymer spectrum \cite{wunder}, while selenium with its $\Delta c_p$ of 1.83 $k_B$ per bead is at the upper end. Obviously, in polymers the structural strain energy is only stored in the three bead degrees of freedom.

\begin{figure}[b]
\hspace{-0cm} \vspace{0cm} \epsfig{file=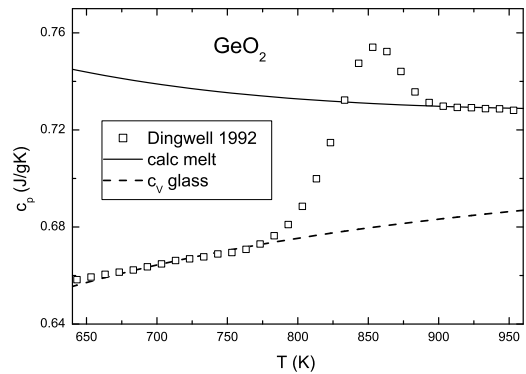,width=7 cm,angle=0} \vspace{0cm} \caption{Heat capacity of glassy and liquid GeO$_2$ \cite{dingwell}. The continuous line is calculated for the liquid as described in the text. The dashed line is determined from the (slightly softened) numerical vibrational density of states \cite{gonzalo}.}
\end{figure}

Fig. 4 shows the measured \cite{dingwell} heat capacity of glassy and liquid GeO$_2$, a network glass former where again all degrees of freedom are expected to contain structural strain energy. These data are measured in heating, which explains the non-equilibrium peak at $T_g$, absent in equilibrium cooling data like those of Fig. 1. The numerically calculated  \cite{gonzalo} glass vibrational density of states, all mode frequencies softened by eight percent, provides the dashed line in Fig. 4. In this case, the contribution from $c_p-c_V$ calculated from the thermal expansion $\alpha_l=7.63$ $10^{-5}$ K$^{-1}$, the liquid bulk modulus $B=8.08$ GPa, and the molar volume 29.13 cm$^3$/mole \cite{dingwell} explain one fifth of $\Delta c_p$, while the rest requires a small temperature-independent free creation energy 0.56 $k_BT_g$, leading to a structural strain contribution of 0.08 $k_B$ per degree of freedom.

Naturally, the question arises: How can GeO$_2$ (and the similar case vitreous silica) freeze into the glassy state with such a low soft mode creation energy? The answer is: In these two cases, the full equilibration by the irreversible viscous shear flow \cite{ab} does not only require a regrouping of local instabilities, as in selenium or in molecular or in metallic glass formers. One actually needs bond breaking processes, adding a strong temperature-independent contribution to the flow barrier.

To summarize, the undercooled liquid looses its structural and vibrational entropy as it approaches the glass temperature by reducing the number of local structural instabilities, which freeze in as boson peak vibrations or low temperature tunneling states at the glass transition. Their density extrapolates to zero at the Kauzmann or Vogel-Fulcher temperature, a concept which contains no paradoxon. The occurrence of structural shear strains in vibrational degrees of freedom with different Gr\"uneisen parameters explains the deviation of the Prigogine-Defay ratio from one.

Very helpful suggestions by Reiner Zorn and Miguel Angel Ramos are gratefully acknowledged.

\end{document}